\documentclass[aps, showpacs]{revtex4-1}
\usepackage{graphicx,amsmath}

\begin{document}

\title{The eigenstate distribution fluctuation from thermal to localized transitions}
\author{Junjun Xu}
\email{Correspondence author. Email: jxu@ustb.edu.cn}
\author{Yanxing Li}
\affiliation{Department of Physics, University of Science and Technology Beijing, Beijing 100083, China}
\date{\today}

\begin{abstract}
We study the thermalization of a quenched disordered Bose-Hubbard model. By considering the eigenstate distribution fluctuation, we show that the thermal to many-body localized transition is always connected to a minimum of this distribution fluctuation. We also observe a Mott-localized regime, where the system fails to thermalize due to the strong on-site repulsion. At last, we show how to detect this eigenstate distribution fluctuation in cold atom systems, which is equivalent to measure the Loschmidt echo of the system. Our work suggests a way to measure the thermal to localized transitions in experiments, especially for a large system.
\end{abstract}

\pacs{72.15.Rn, 73.20.Jc, 05.30.−d}
\maketitle

Recently there has been extensive progress in the study of non-equilibrium dynamics of a closed quantum many-body system, for example, the experimental observation of a dynamical quantum phase transition~\cite{Roos, Sengstock, Heyl} and the thermal to many-body localized (MBL) transition~\cite{Schreiber, Choi, Smith, Bordia1, Bordia2, Luschen1, Luschen2, Nandkishore, Altman, Abanin, Alet}, both in trapped ions and cold atom systems. The realization of these phenomena in experiments enrich our knowledge between quantum mechanics and equilibrium statistical mechanics. We will study the thermalization in an optically trapped Bose gas, motivated by the recent experiment carried out in Greiner's group, and show how the thermalization and many-body localization are connected to the information of its initial state.

Typically, a non-equilibrium closed quantum system can be described by a single pure state $|\psi(t)\rangle$, which undergoes a unitary evolution as $|\psi(t)\rangle=e^{-iHt}|\psi(0)\rangle$ according to the Schr\"odinger equation, where $H$ is the Hamiltonian of the closed system. If we project this state into its eigenstates,  with $|\psi(t)\rangle=\sum_me^{-iE_mt}c_m|\psi_m\rangle$, where $m$ labels the $m$-th eigenstate, we can find there is always a memory of the initial eigenstate distribution $p_m=|c_m|^2$ in its dynamics. This is distinct from our knowledge of quantum statistical mechanics, where an equilibrium closed quantum system should have an equal probability of its allowed states under the micro-canonical ensemble. However, recent studies have revealed the connection between these two cases, where a long-time averaging of the local observables in the non-equilibrium dynamics is found well predicted by the thermal ensemble, while the system's initial information is hidden locally. Such a relaxed non-equilibrium many-body state is called a thermalized state. Deutsch first considers the problem and shows by adding a random matrix perturbation, local observables of a closed quantum system agree with the micro-canonical distribution~\cite{Deutsch}. Srednicki further shows that in some particular systems, such as a dilute hard-sphere gas, the eigenstate itself obeys the thermal distribution for the momentum of each particle, which is termed as the `eigenstate thermalization'~\cite{Srednicki}. Such behavior has been observed experimentally in cold atom systems~\cite{Rigol, Deutsch2}.

In fact, not all non-equilibrium closed quantum systems will thermalize. An interesting example is the interacting lattice Bose or Fermi gases with a strong local disorder. Experiments show that this system can stay in its initial state for a sufficiently long time, the phenomena which is called the many-body localization~\cite{Nandkishore, Altman, Abanin, Alet}. To understand why and when these systems fail to thermalize and find the characteristic behaviors of the thermal to localized transitions have become a major goal in this field, and extensive works have been carried out recently. For example, one finds the MBL state can be distinguished from the thermal state by an unbounded logarithmic entanglement growth\cite{Znidaric, Bardarson, Serbyn2, Vosk2}. Researches have shown recently there is a rare-region Griffiths regime near the thermal to MBl transition, with a slow dynamics~\cite{Griffiths, Agarwal1, Gopalakrishnan1, Vosk, Gopalakrishnan2, Luitz, Agarwal2}. This transition is also found connected to the level spacing statistics, with the thermalized state shows strong level repulsion and thus a Wigner-Dyson distribution, while the MBL state shows a Poisson distribution~\cite{Oganesyan, Pal, Iyer, Modak, Ponte, Serbyn, Alessio, Sierant}.

In this letter, we consider the recent experiment carried out by Greiner's group~\cite{Greiner18}. They prepare an array of one-dimensional $^{87}{\rm Rb}$ atoms in a deeply optical lattice. The non-equilibrium dynamics of their system is driven by first preparing the atoms in a Mott state with one atom per site, then quenching the lattice depth to a shallow value, which opens up the hoping of atoms between sites. The information of the time evolution is given by a series of single-site number-resolved imaging~\cite{Bakr, Kuhr, Kaufman}. In their experiment, as the lattice depth is larger than the optical recoil energy, the system is well described by the following disordered Bose-Hubbard model
\begin{eqnarray}
H=&&-J\sum_i\left(\hat{a}_i^\dagger \hat{a}_{i+1}+{\rm H.c.}\right)+\frac{U}{2}\sum_{i}\hat{n}_i\left(\hat{n}_i-1\right)\nonumber\\
&&+\sum_ih_i\hat{n}_i,
\end{eqnarray}
where $J$ and $U$ account for the nearest hopping and on-site interaction, with $\hat{n}_i=\hat{a}_i^\dagger \hat{a}_i$ is the particle number operator on site $i$. The disorder potential is chosen as the Aubry-Andr\`e model $h_i=W\cos(2\pi\beta i+\phi)$, with $\beta=(\sqrt{5}+1)/2$ an irrational number, $\phi$ a random phase, and $W$ gives the strength of this disorder. This form of disorder has been realized in various experimental groups~\cite{Schreiber, Bordia1, Bordia2, Luschen1, Luschen2, Aubry, Roati}.

\begin{figure}[h]
 \includegraphics[width=0.45\textwidth]{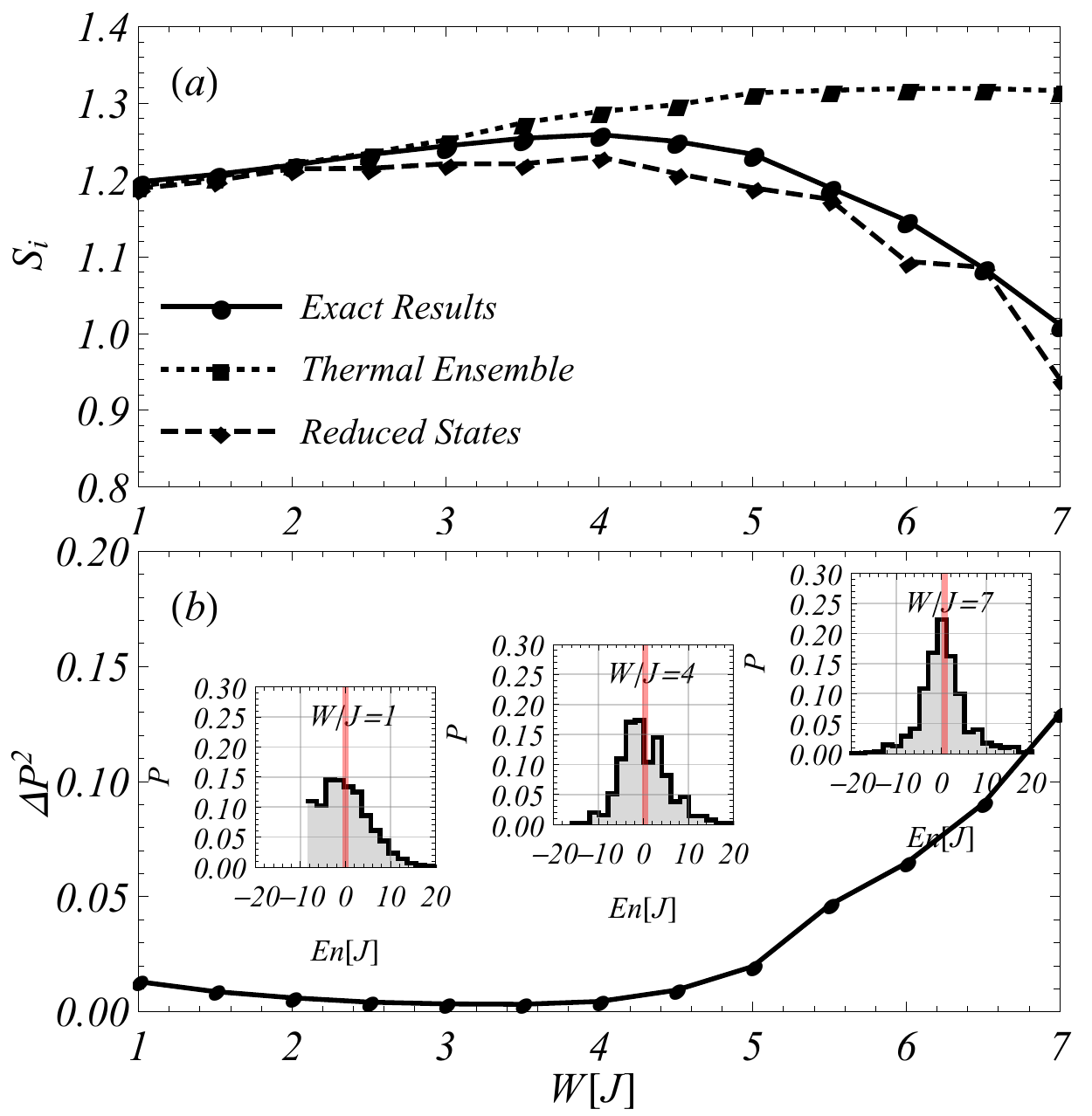}
  \caption{(a) The single-site entanglement entropy $S_i$ (time-averaged from $t=10^2/J$ to $t=10^4/J$) of the disordered Bose-Hubbard model for different disorder strength $W$. The solid line is our exact results and the dotted line is the thermal prediction assuming a micro-canonical ensemble of nearest 11 eigenstates around the energy of the system $E_0=\langle\psi(0)|H|\psi(0)\rangle$. The dashed line shows the exact dynamics of these nearest 11 eigenstates, the results of which is not sufficient to describe the thermal and localized states. (b) The eigenstate distribution fluctuation $\Delta P^2=\sum_m(p_m-\bar{p}_m)^2$ of the whole Hilbert space for different disorder strength $W$. The inset shows the corresponding eigenstate distribution $P$ for an initial Mott state, with the vertical red line indicates the energy of the system. The calculation is taken for a chain length of $L=8$ with $N=8$ and $U/J=2.87$.}
  \label{fig:fig1}
\end{figure}

We are aiming to find an appropriate way to locate the thermal to localized transitions in this work. As the experiment, we quench the system from a Mott state. Since this is a relatively closed system, such quenching dynamics will not relax to a thermal equilibrium state. Previous studies have shown that such Aubry-Andr\'e like disorder will support a thermal to MBL transition. For example, one can measure the half entanglement entropy, which will show an unbounded logarithmic increase in the MBL state. Such findings, however, are hard to observe in the experiments, due to the large configuration space of this highly-excited system. For $8$ bosons on a chain length of $L=8$, the configuration space of the half system is about $(N+1)^{L/2-1}=729$, which shows an exponential increase as a function of the chain length. To measure this half entanglement entropy, one needs to project the results to each configuration, which is far beyond the experimental realization.

An alternative is to do a local observation. For example, instead of measuring the half entanglement entropy, one can measure the single-site entanglement entropy. This measurement has been carried out recently by Greiner's group~\cite{Greiner18}. The similar results have been reproduced in Fig.~\ref{fig:fig1}(a) as the solid and dotted lines. They find the deviation between the single-site entanglement entropy and the thermal prediction increases as one increase the disorder strength, signaling a thermal to MBL transition.

In the work, we use both a time-dependent matrix product state (MPS) algorithm (implemented in the ALPS package~\cite{alps, alpsdetail}) and an exact diagonalization calculation to study these non-equilibrium dynamics, the results of which coincide with each other. Throughout this paper, we run a calculation with $N=8$ bosons on a chain length of $L=8$ with the maximum matrix dimension $6435\times6435$. Due to this huge Hilbert space, it's beyond our current ability to push this calculation above $L=8$. The disorder averages are over 20 different realizations. We confirm there is no significant difference if we add more disorder realizations. 

Our single-site entanglement entropy results are illustrated in Fig.~\ref{fig:fig1}(a) with $S_i=- \mathrm{Tr}(\hat{\rho}_i\log\hat{\rho}_i)$, where $\hat{\rho_i}=\mathrm{Tr}^i_{L-1}\hat{\rho}=\mathrm{Tr}^i_{L-1}|\psi\rangle\langle\psi|$ is the reduced density matrix at site $i$ with $|\psi\rangle$ the state of the system and the trace is over the remaining $L-1$ sites. To avoid edge effects, we neglect the two edge sites and the entropy $S_i$ in Fig.~\ref{fig:fig1}(a) is the average over the middle 6 sites. The thermal prediction is calculated assuming a micro-canonical ensemble of the nearest $l$ eigenstates around the energy of the system $E_0=\langle\psi(0)|H|\psi(0)\rangle$, with $\hat{\rho}=1/l\sum_{m=1}^l|\psi_m\rangle\langle\psi_m|$ the thermal density matrix of the system and $|\psi_m\rangle$ is the $m$-th eigenstate of the system. Thus in the thermal prediction, the single-site density matrix is $\hat{\rho}_i=1/l\sum_{m=1}^l\hat{\rho}_i^m$, with $\hat{\rho}_i^m=\mathrm{Tr}^i_{L-1}|\psi_m\rangle\langle\psi_m|$ the reduced single-site density matrix of the $m$-th eigenstate. As in~\cite{Greiner18}, we choose $l=11$ and we confirm there is no significant difference if we shift $l$ in a reasonable range.

To further explore this thermal to MBL transition, we look at the exact dynamics starting from the initial state including only these 11 nearest eigenstates around the system energy $E_0=\langle\psi(0)|H|\psi(0)\rangle$, which is shown as the dashed line in Fig.~\ref{fig:fig1}(a). Since these 11 eigenstates will be sufficient to give an accurate prediction in the thermalized regime as the dotted line in Fig.~\ref{fig:fig1}(a) for $W<3J$, one may assume that this system is determined by only these nearest eigenstates, and these states have small distribution fluctuations which lead to the micro-canonical ensemble. However, this assumption is false. From the dashed line in Fig.~\ref{fig:fig1}(a), one can see these states fail to give a good prediction at around $W=3J$. One can also calculate the fluctuation of these nearest eigenstates, which is found irregular in the whole regime. This means the thermalization, even though it can be accurately predicted by its nearest eigenstates, depends on the total eigenstate space.

Then how about the fluctuation of the total eigenstates distribution? We check this fluctuation $\Delta P^2=\sum_m(p_m-\bar{p}_m)^2$ in Fig.~\ref{fig:fig1}(b), with $p_m$ the probability of the initial state in its $m$-th eigenstate. The results show that this fluctuation is relatively small in the thermalized regime, while it becomes large in the MBL state. In the inset of Fig.~\ref{fig:fig1}(b), we show the eigenstate distribution for different disorder strength. For large disorder $W=7J$, the distribution is shown localized around the total energy and has a large distribution fluctuation.

\begin{figure}[h]
 \includegraphics[width=0.38\textwidth]{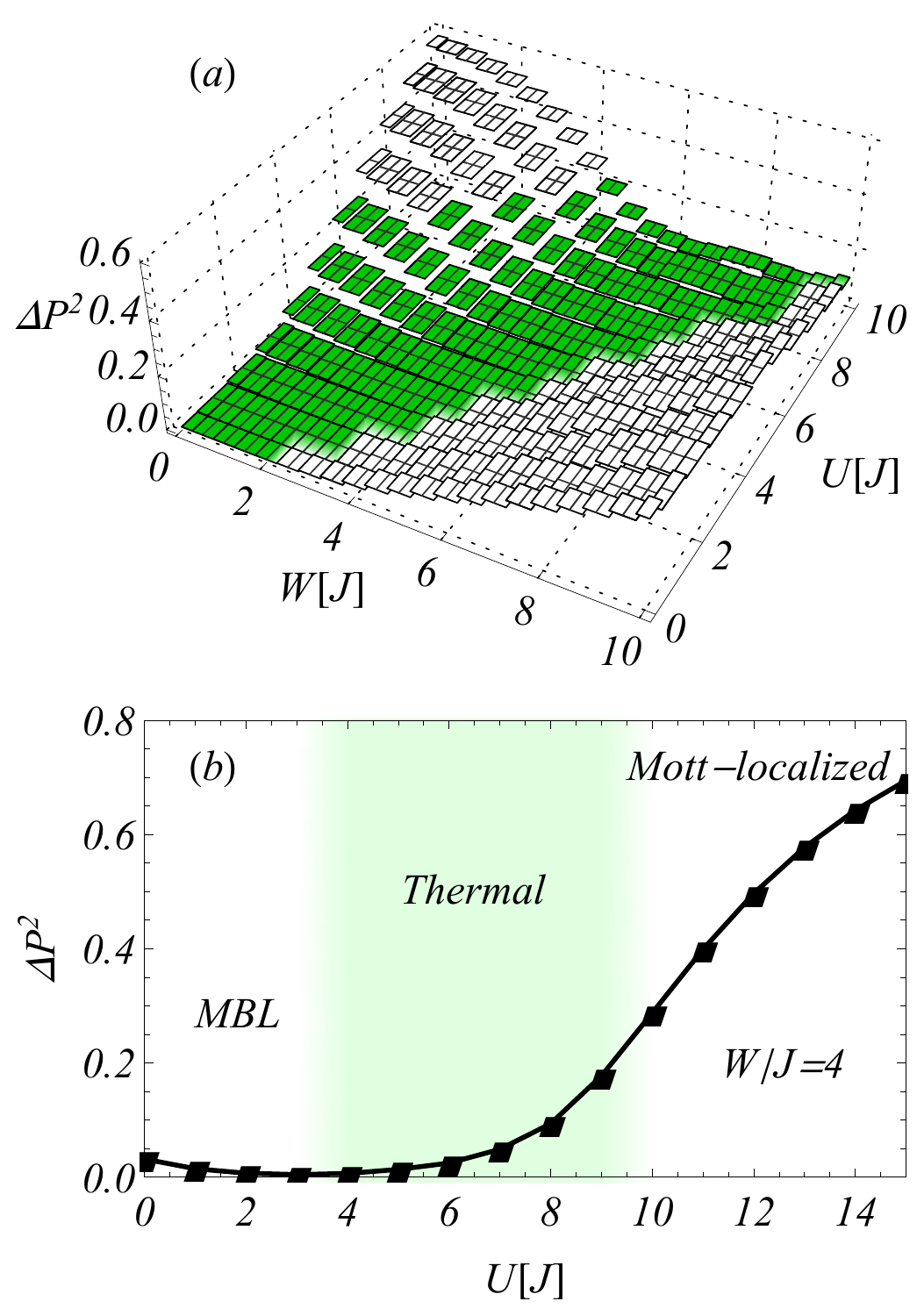}
  \caption{(a) The calculated eigenstate distribution fluctuation $\Delta P^2$ for different disorder and interaction strength $W$ and $U$. The green color represents the regime where the thermal ensemble gives a good prediction of the equilibrium single-site entanglement entropy. There are two regimes that break this thermalization, which is the MBL and Mott-localized regime, respectively. (b) A typical illustration of this thermal to localized transitions. The thermal to MBL transition is driven by a strong disorder and happens around the minimum of the eigenstate distribution fluctuation, with a weaker interaction $U$. At large $U$, the fail of thermalization is due to the strong on-site repulsion, which suppresses the growth of entanglement entropy in this system.}
  \label{fig:fig2}
\end{figure}

\begin{figure}[h]
 \includegraphics[width=0.45\textwidth]{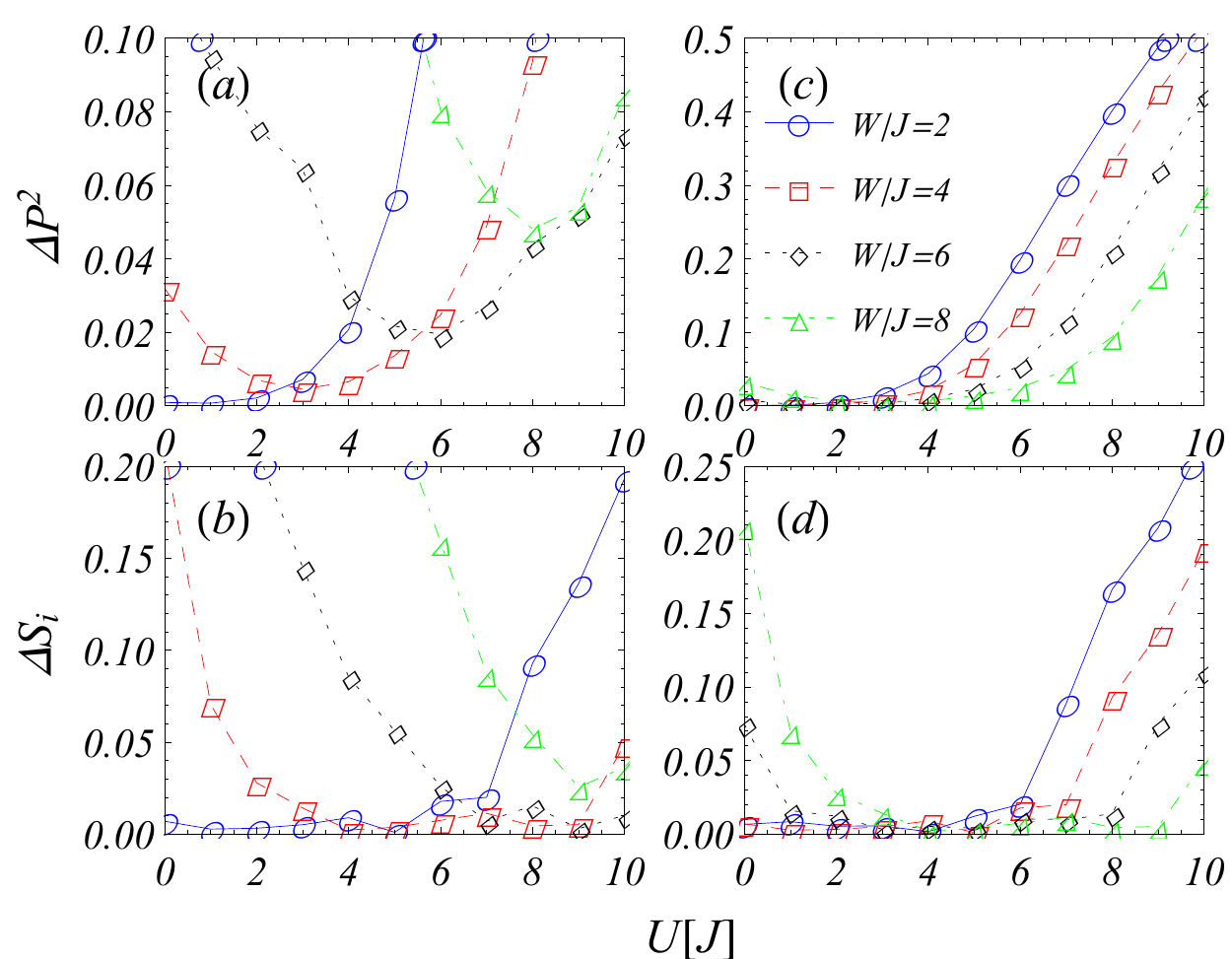}
  \caption{An illustration how the eigenstate distribution fluctuation $\Delta P^2$ is connected to the thermal to localized transitions. (a) and (c) show the fluctuations $\Delta P^2$ as a function of interaction strength $U$ for various disorder $W/J=2,4,6,8$. (b) and (d) show the corresponding deviation of the calculated single-site entanglement entropy $\Delta S_i$ from the thermal prediction. From the left figures (a) and (b), there is always a minimum of $\Delta P^2$ where the entropy deviation $\Delta S_i$ starts to show a obvious increase. Thus we attribute this minimum of $\Delta P^2$ as a signature of a thermal to MBL transition. The right two figures (c) and (d) illustrate how the on-site localization breaks the thermalization.}
  \label{fig:fig3}
\end{figure}

In Fig.~\ref{fig:fig2}(a) we show the eigenstate distribution fluctuations for different disorder strength $W$ and interaction strength $U$. The green area labels the thermalized regime, which is determined by calculating the deviation of the exact and thermal predicted single-site entanglement entropy. Typically, for large enough disorder or interaction, one would see a localization of the many-body wavefunctions, leading to an increase of the distribution fluctuations $\Delta P^2$ as shown in Fig.~\ref{fig:fig2}(a). The former case is an MBL state, while the latter one is termed here as the Mott-localized state. In Fig.~\ref{fig:fig2}(b), we show these two thermal to localized transitions as a function of interaction strength $U$ for a typical disorder $W=4J$. Even though these two localized states both happen at finite disorder and interaction here, the mechanisms are different: the MBL state is due to the localization at some local disorder potentials, while the Mott-localized state is due to the strong on-site repulsion that giving rise to the on-site localization. We note in the Mott-localized state, the entanglement entropy is highly suppressed, and there is no logarithmic growth of the half entanglement entropy as the MBL state. 

Figure~\ref{fig:fig3} shows the details how this eigenstate distribution fluctuation $\Delta P^2$ is connected to the thermal to localized transitions. In Fig.~\ref{fig:fig3}(a)-(b) the deviation of single-site entanglement entropy from our thermal prediction $\Delta S_i$ always shows an obvious increase around the minimum of $\Delta P^2$. Thus we locate the thermal to MBL transition at the minimum of  $\Delta P^2$. The thermal to Mott-localized transition has been previously considered by Kollath {\it et al.}, and is found connected to the quasiparticle interactions in the Mott regime \cite{Kollath}. We show such transition in Fig.~\ref{fig:fig3}(c)-(d), and we find a monotonous increase of the distribution fluctuation as we increase the on-site interaction. Thus it is hard to locate this transition. We consider the thermal prediction to be good for $\Delta S_i<0.1$, and then we could illustrate the boundary of this thermal regime at $\Delta P^2\approx0.25$.

We now come to the experimental measurement of such eigenstate distribution fluctuation. In fact, it is related to the Loschmidt echo in the experiment. If we prepare the system at its eigenstate and then at time $t=0$ quench it to a new Hamiltonian $H$, the Loschmidt echo (which is also termed as the survival probability) measures the probability how much the system belongs to the initial state as
\begin{equation}
L(t)=\left|\langle\psi(0)|e^{-iHt}|\psi(0)\rangle\right|^2.
\end{equation}
Recent studies have shown its importance in describing the non-equilibrium dynamics of a closed quantum many-body system. For example, a nonanalytic kink in this Loschmidt echo is found connected to a dynamical phase transition~\cite{Heyl}. The Loschmidt echo of the MBL state has been considered previously by Torres-Herrera and Santos in a 1D isotropic Heisenberg spin-1/2 model, and they find it has a power-law decay at long times, which corresponds to the multifractal states of the system~\cite{Santos}.

\begin{figure}[h]
 \includegraphics[width=0.48\textwidth]{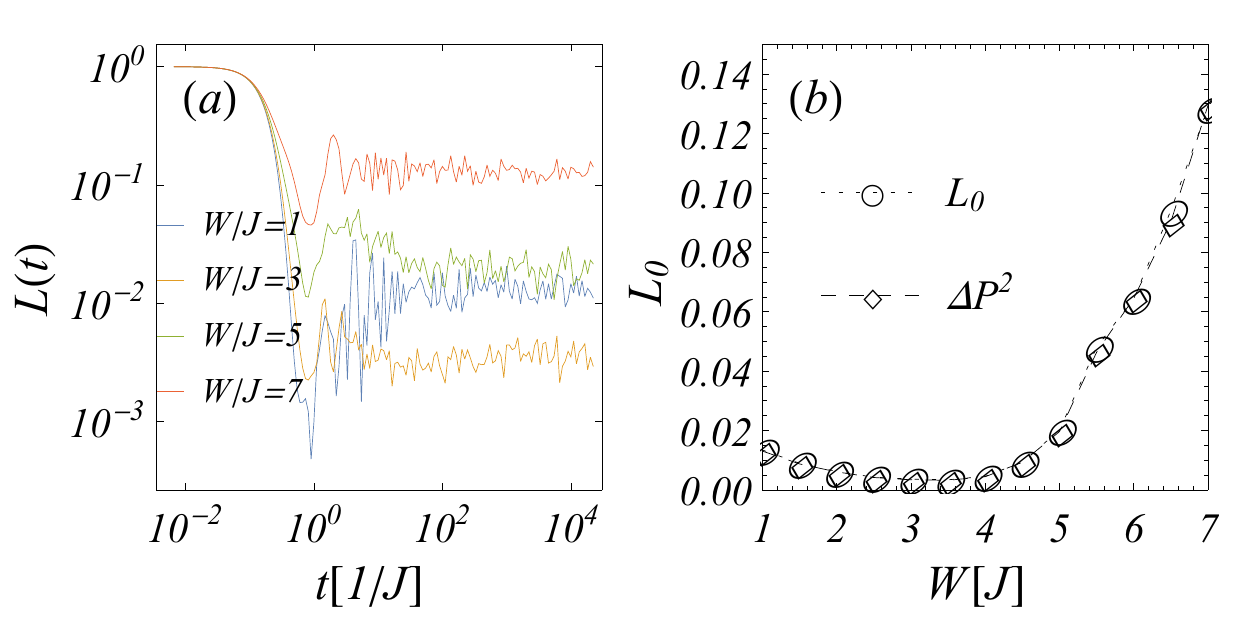}
  \caption{(a) The Loshmidit echo $L(t)$ as a function of evolution time $t$ for different disorder strength $W$. (b) The time-averaged Loschmidt echo $L_0$ for different disorder strength $W$. For comparison, we also show the eigenstate distribution fluctuation $\Delta P^2$ as the dashed line in (b), which is almost the same with the averaged $L_0$. The calculation is taken for a chain length of $L=8$ with $N=8$ and $U/J=2.87$.}
  \label{fig:fig4}
\end{figure}

Here we show the Loschmidt echo for different disorder strength in Fig.~\ref{fig:fig4}(a). After sufficient long period of time, the system relaxes to an equilibrium state and the Loschmidt echo $L(t)$ reaches a stationary value, due to the many-body dephasing. Similar as Fig.~\ref{fig:fig1}(b), we see a non-monotonic behavior of the stationary value as we increase the disorder strength. To check how these stationary values change, we plot the time-averaged Loschmidt echo $L_0$ from $t=10^2/J$ to $t=10^4/J$ as a function of disorder strength $W$ in Fig.~\ref{fig:fig4}(b). For comparing, we also show the eigenstate distribution fluctuation $\Delta P^2$ as the dashed line. The difference between these two lines are negligible, suggesting these two measurements are equivalent. This can be understood as following. If we expand the initial state in the eigenstate space as $|\psi(0)\rangle=\sum_m c_m|\psi_m\rangle$, the Loschmidt echo can be written as
\begin{eqnarray}
L(t)&&=\sum_{mn}|c_m|^2|c_n|^2e^{-i(E_m-E_n)t}\nonumber\\
&&=\sum_m|c_m|^4+\sum_{m>n}2\cos\left[\left(E_m-E_n\right)t\right]|c_m|^2|c_n|^2.\nonumber
\end{eqnarray}
Thus the time evolution of this Loschmidt echo includes an oscillating off-diagonal part, which becomes less important when the long-time dephasing is included. The time averaged Loschmidt echo $L_0$ will goes to the diagonal part. Considering our eigenstate fluctuation $\Delta P^2=\sum_m(p_m-\bar{p})^2=\sum_m|c_m|^4-1/d$ with $d$ the dimension of Hilbert space. Since $d$ is very large for a typical system, the time-averaged Loschmidt echo $L_0$ is almost always predict the eigenstate distribution fluctuation $\Delta P^2$.

The experimental observation of such Loschmidt echo is free of large configuration space and accessible in current cold atoms high-resolution single-site number-resolved imaging technique~\cite{Bakr, Kuhr, Kaufman}. Comparing with the entanglement detection, one does not need to count each number configuration, instead one just need to pick the Mott-insulating state and find the probability of the final state in this configuration space. Thus one may push the experiments to a larger system size, which will be more helpful for the studying of the intermediate Griffiths regions and the thermal to localized transitions at thermodynamic limit.

To summarize, we have proposed a way to measure the thermal to localized transitions in the disordered Bose-Hubbard model, by looking at their eigenstate distribution fluctuations. We find the thermal to MBL transition is always connected to a minimum of the eigenstate distribution fluctuation. We also observe a Mott-localized regime, where the system is localized due to the strong on-site repulsion. We then show how this fluctuation is equivalent to the time-averaged Loschmidt echo, which is accessible in current cold atoms experiments. We note at last the method of measuring the Loschmidt echo here is directly applicable to trapped-ion systems~\cite{Heyl, Smith}, and maybe also feasible in solid state systems where a critical thermalization dynamics is observed recently~\cite{Kucsko}.

We acknowledge Markus Greiner and Matthew Rispoli for sharing their experiment details. This research is supported by NNSFC (No. 11504021) and FRFCU (No. FRF-TP-17-023A2).


\begin{thebibliography}{10}
\bibitem{alpsdetail} The single-site entanglement entropy is get by calculating the number probabilities on site $i$, and in ALPS it is equivalent to calculate the expectation value of the projection operator $\hat{p}_{n_i}=\left[\prod_{m=1,m\neq n_i}^N(\hat{n}-m)\right]/\left[\prod_{m=1,m\neq n_i}^N(n_i-m)\right]$
\end{thebibliography}
\end{document}